\documentclass[12pt]{article}
\usepackage{graphicx}

\newcommand\pubnumber{NuPhys2016-Hallsj\"o}
\newcommand\pubdate{\today}

\def\support{\footnote{This project has received funding from the European Union's Horizon 2020 Research and Innovation programme under grant agreement No. 654168.}}

\renewcommand{\thefootnote}{$\star$}

\def\Title#1{\begin{center} {\Large #1 } \end{center}}
\def\Author#1{\begin{center}{ \sc #1} \end{center}}
\def\Address#1{\begin{center}{ \small{\it #1}} \end{center}}

\newcommand\pubblock{\rightline{\begin{tabular}{l} \pubnumber\\
         \pubdate  \end{tabular}}}
\newenvironment{Abstract}{\begin{quotation}  }{\end{quotation}}
\newenvironment{Presented}{\begin{quotation} \begin{center} 
             PRESENTED AT\end{center}\bigskip 
      \begin{center}\begin{large}}{\end{large}\end{center} \end{quotation}}

\def\beq{\begin{equation}}
\def\eeq#1{\label{#1}\end{equation}}
\def\eeqn{\end{equation}}

\def\beqa{\begin{eqnarray}}
\def\eeqa#1{\label{#1}\end{eqnarray}}
\def\eeqan{\end{eqnarray}}

\let\bar=\overbar

\def\Dslash{\not{\hbox{\kern-4pt $D$}}}
\def\dslash{\not{\hbox{\kern-2pt $\del$}}}

\def\msb{{\bar{\ssstyle M \kern -1pt S}}}

\begin{document}
\begin{titlepage}
\pubblock

\vfill
\Title{Baby MIND: A magnetised spectrometer for the WAGASCI experiment}
\vfill

\Author{M.~Antonova$^{1}$, R.~Asfandiyarov$^{2}$, R.~Bayes$^3$, P.~Benoit$^4$,  A.~Blondel$^2$, M.~Bogomilov$^5$, A.~Bross$^6$, F.~Cadoux$^2$, A.~Cervera$^{7}$, N.~Chikuma$^{8}$, A.~Dudarev$^4$, T.~Ekel\"of$^9$, Y.~Favre$^2$, S.~Fedotov$^1$,  S-P.~Hallsj\"o$^{3}$, A.~Izmaylov$^1$ , Y.~Karadzhov$^2$, M.~Khabibullin$^{1}$, A.~Khotyantsev$^1$, A.~Kleymenova$^{1}$, T.~Koga$^{8}$, A.~Kostin$^{1}$, Y.~Kudenko$^{1}$, V.~Likhacheva$^{1}$, B.~Martinez$^2$, R.~Matev$^5$, M.~Medvedeva$^{1}$, A.~Mefodiev$^{1}$, A.~Minamino$^{10}$, O.~Mineev$^1$, M.~Nessi$^4$, L.~Nicola$^2$, E.~Noah$^2$, T.~Ovsiannikova$^{1}$, H.~Pais Da Silva$^4$, S.~Parsa$^2$, M.~Rayner$^4$, G.~Rolando$^4$, A.~Shaykhiev$^1$, P.~Simion$^9$, F.J.P.~Soler$^3$, S.~Suvorov$^{1}$, R.~Tsenov$^5$, H.~Ten Kate$^4$, G.~Vankova-Kirilova$^5$ and N.~Yershov$^1$.\support}


\Address{$^1$Institute of Nuclear Research, Russian Academy of Sciences, Moscow, Russia.\\
$^2$University of Geneva, Section de Physique, DPNC, Geneva, Switzerland.\\
$^3$University of Glasgow, School of Physics and Astronomy, Glasgow, UK.\\
$^4$European Organization for Nuclear Research, CERN, Geneva, Switzerland.\\
$^5$University of Sofia, Department of Physics, Sofia, Bulgaria.\\
$^6$Fermi National Accelerator Laboratory, Batavia, Illinois, USA.\\
$^{7}$IFIC (CSIC $\&$ University of Valencia), Valencia, Spain.\\
$^8$University of Tokyo, Tokyo, Japan.\\
$^9$University of Uppsala, Uppsala, Sweden.\\
$^{10}$Yokohama National University, Yokohama, Japan.
}

\vfill
\begin{Abstract}
The WAGASCI experiment being built at the J-PARC neutrino beam line will measure the difference in cross sections from neutrinos interacting with a water and scintillator targets, in order to constrain neutrino cross sections, essential for the T2K neutrino oscillation measurements. A prototype Magnetised Iron Neutrino Detector (MIND), called Baby MIND, is being constructed at CERN to act as a magnetic spectrometer behind the main WAGASCI target to be able to measure the charge and momentum of the outgoing muon from neutrino charged current interactions.
\end{Abstract}
\vfill
\begin{Presented}
NuPhys2016, Prospects in Neutrino Physics\\
Barbican Centre, London, UK,  December 12--14, 2016
\end{Presented}
\vfill
\end{titlepage}
\def\thefootnote{\fnsymbol{footnote}}
\setcounter{footnote}{0}

\section{Introduction}

The prototype Magnetized Iron Neutrino Detector (Baby MIND) \cite{NP05} is currently being built at CERN, where it will be fully characterized at a charged particle beam line. Once the detector has been characterised, the plan is to integrate it into the WAGASCI, WAter Grid And SCIntillator, (T59) experiment \cite{Koga:2015iqa,Noah:2015jgd}  in Japan to improve measurements of the ratio of neutrino interaction cross sections on water and carbon. To this end, it will provide charge identification and momentum measurements for muons resulting from neutrino interactions in the WAGASCI neutrino targets located upstream. The purpose of this experiment is to constrain the main non-cancelling systematic uncertainty for the neutrino oscillation analysis at T2K \cite{Abe:2013hdq}.

\section{WAGASCI and Baby MIND layout}

WAGASCI consists of a segmented target of water and scintillator cells, where the cross section can be measured in both media simultaneously. The WAGASCI experiment requires some form of magnetic spectrometer to measure momentum and charge identification of the outgoing muons from charged current interactions. The Baby MIND detector consists of 33 magnetised metal plates and 18 scintillator modules that measure the position of hits along the spectrometer and the curvature of the track in the magnetic field. Two muon range detectors (MRD) are placed at either side of the WAGASCI target to measure escaping particles.  

\begin{figure}[h]
		\begin{minipage}{0.45\linewidth}
			\centerline{\includegraphics[width=0.9\linewidth]{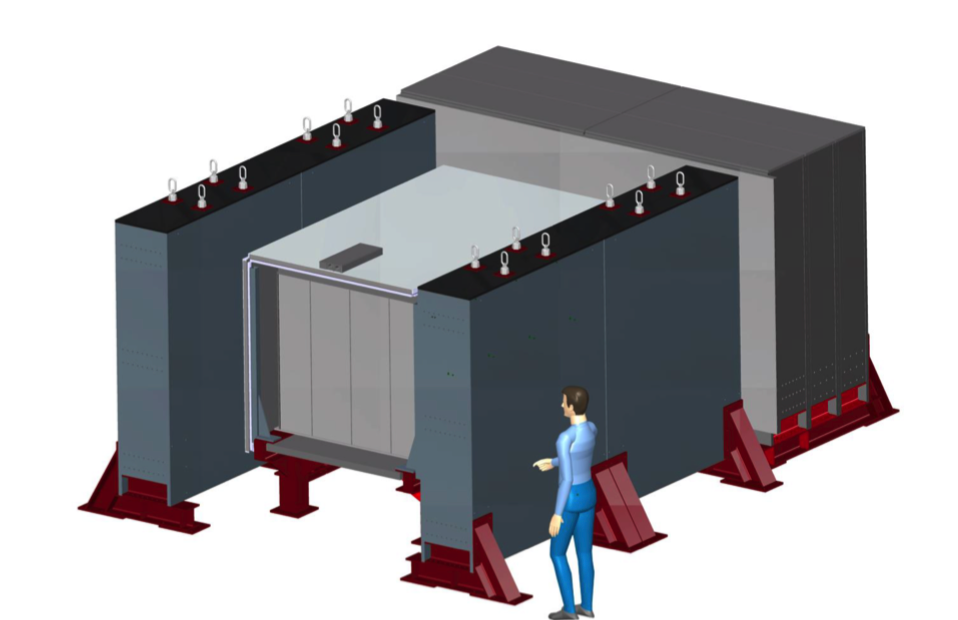}}
			\caption{Baby MIND integrated into the WAGASCI experiment.}
			\label{fig:MINDWagasci}
		\end{minipage}
		\hfill
	\begin{minipage}{0.51\linewidth}
		\centerline{\includegraphics[width=0.9\linewidth]{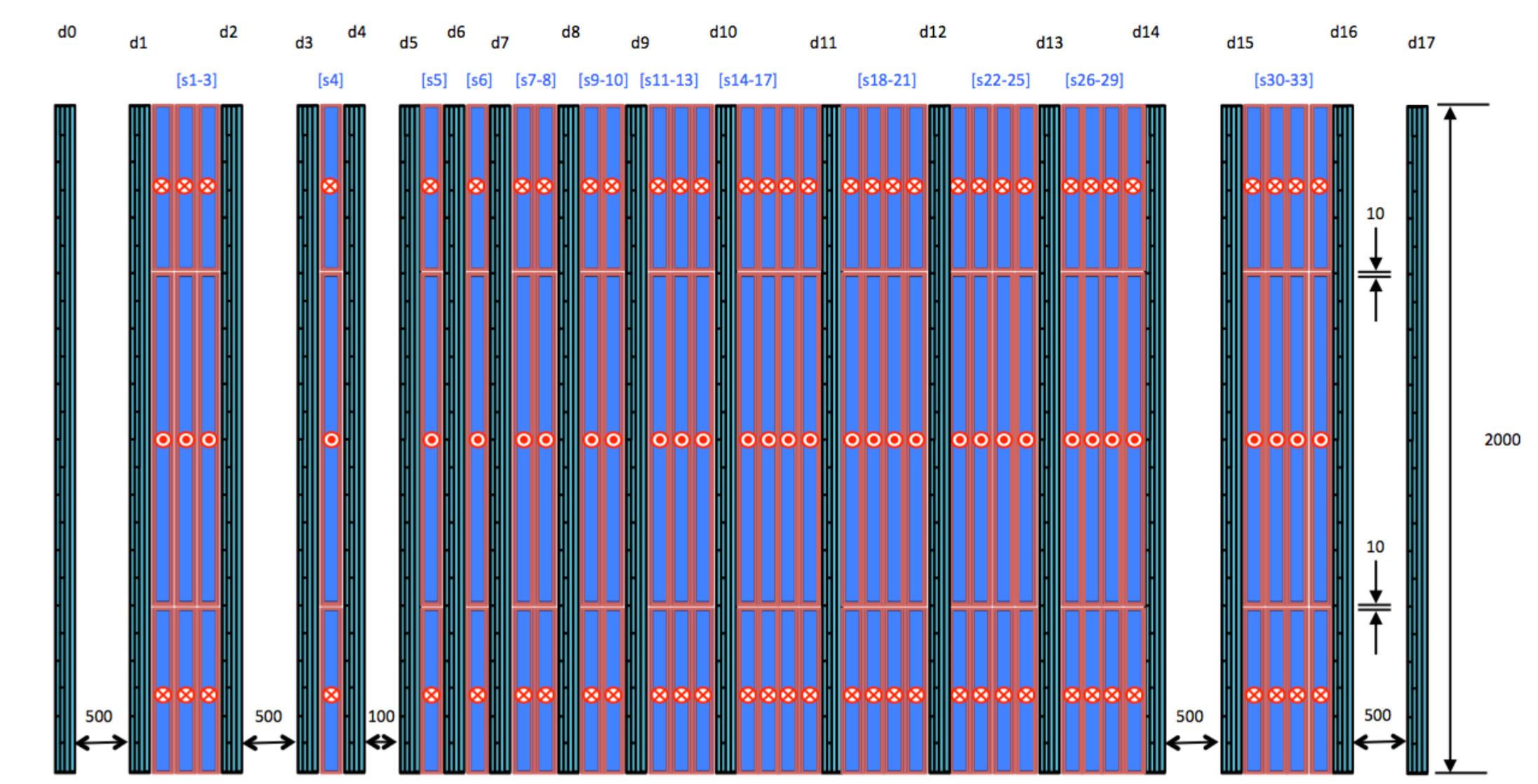}}
			\caption[]{A side view of Baby MIND with scintillator planes (grey) and magnetised iron (blue/red).}
				\label{fig:babyMIND}
	\end{minipage}
\end{figure}

Figure~\ref{fig:MINDWagasci} shows a layout of the WAGASCI detector, with the main target at the centre, Baby MIND at the end and the two side-MRDs. Figure~\ref{fig:babyMIND} shows a more detailed side view of the Baby MIND detector, with the alternating magnetised iron plates and scintillator planes shown. 
Each scintillator module consist of four planes of polysterene-based extruded scintillator bars,  two of the planes are oriented along the horizontal direction, with bars 30~mm wide, and two of the planes are oriented along the vertical direction, with bars 210~mm wide. Each bar contains Kuraray wavelength-shifting fibres of diameter 1.0~mm to collect the light and transport it at either end to Hamamatsu S12571-025C Multi-pixel Photon Counters (MPPC).

\section{Electronics test beam}
\begin{figure}[h]
	\begin{minipage}{0.39\linewidth}
		\centerline{\includegraphics[width=0.9\linewidth]{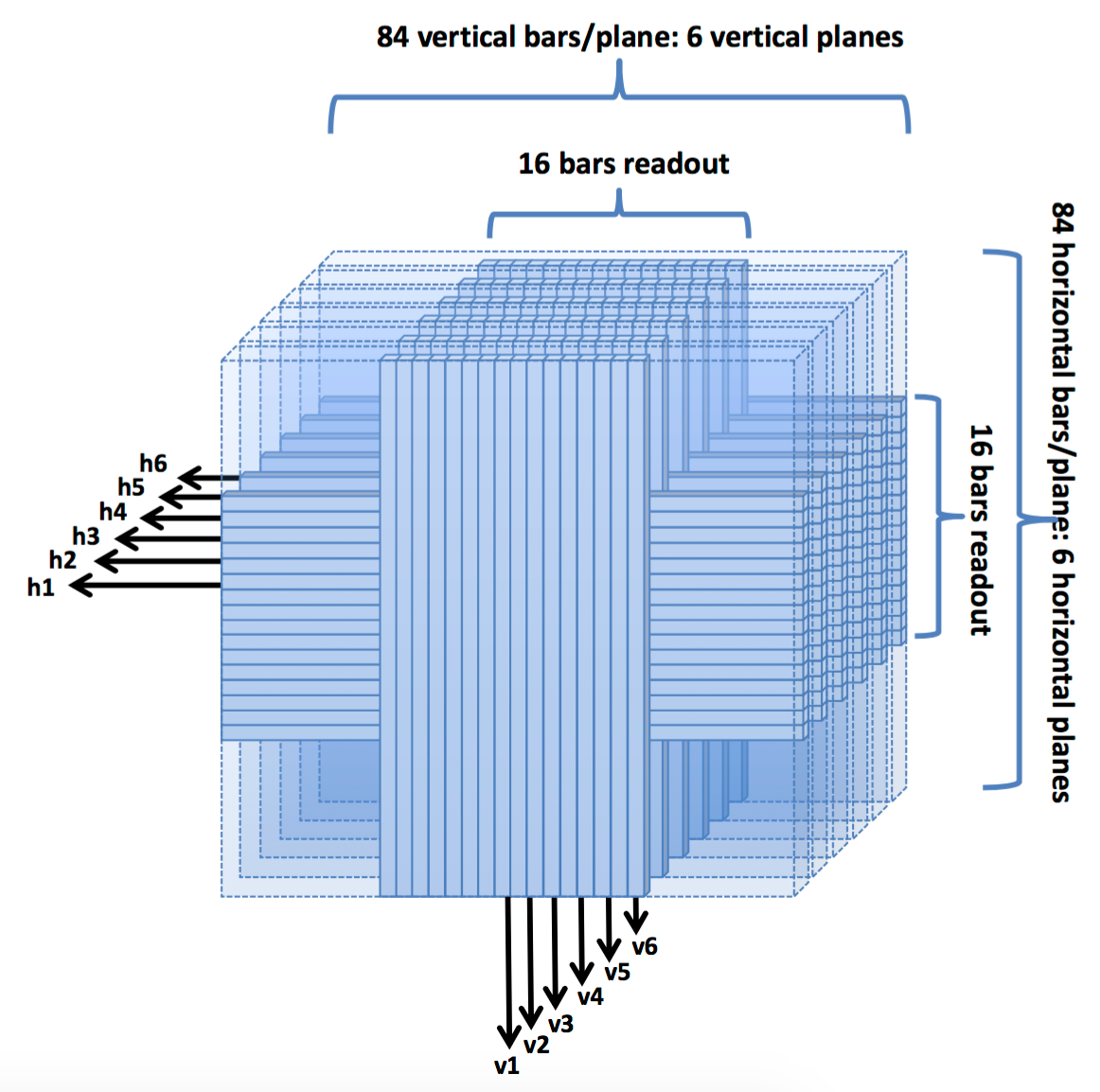}}
		\caption[]{The TASD with the instrumented bars visualised.}
		\label{fig:AIDA}
	\end{minipage}
	\hfill
	\begin{minipage}{0.59\linewidth}
		\centerline{\includegraphics[width=\textwidth, trim = 5cm 5cm 5cm 5cm]{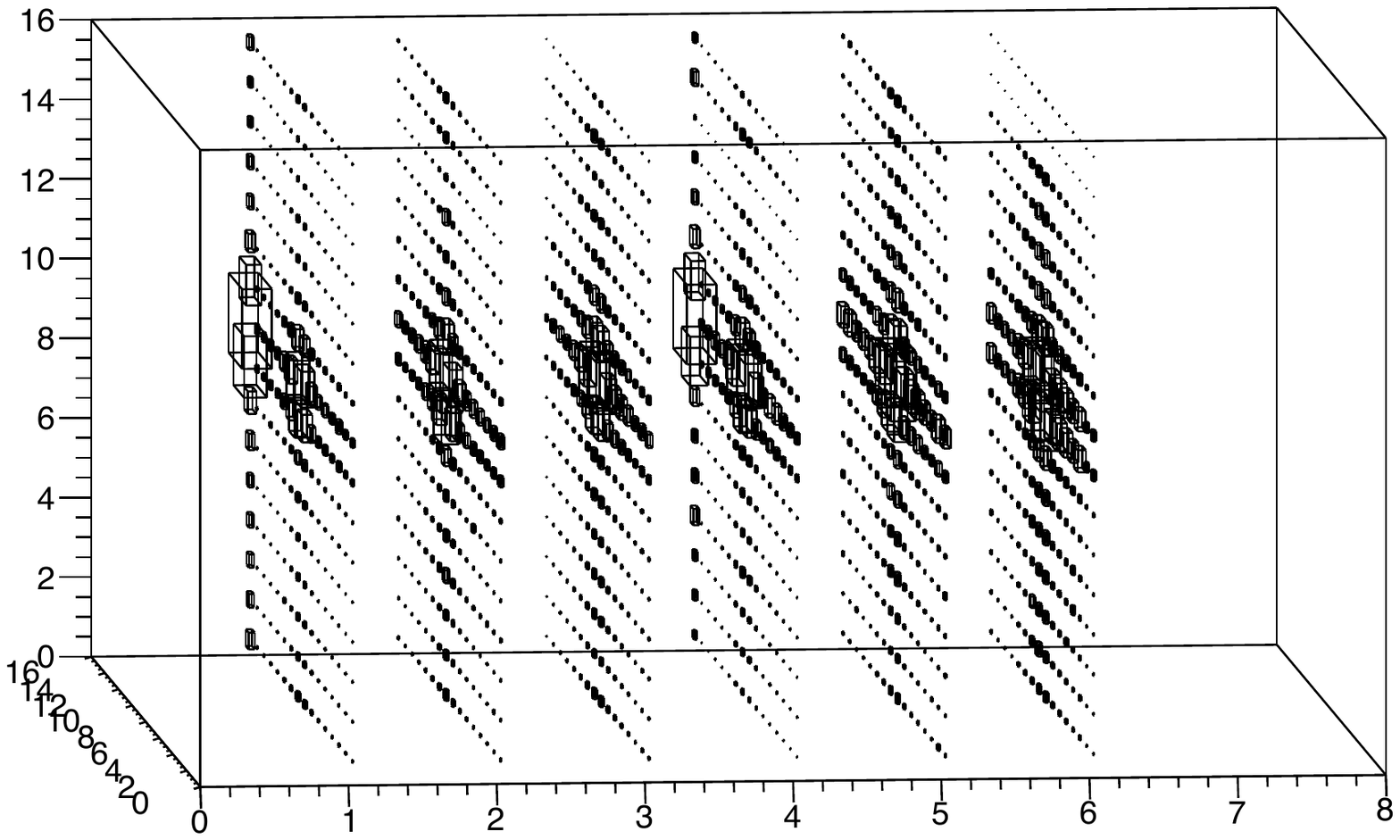}}
		\caption[]{Beam profile measured by the TASD.}
		\label{fig:beamProfile}
	\end{minipage}
\end{figure}

During June-July 2016 a test beam was performed to characterise the readout system, data acquisition (DAQ) and electronics to be used in the Baby MIND detector. The test beam was at the T9 beam of the East Area, operating at the Proton Synchrotron (PS) at CERN. A Totally Active Scintillation Detector (TASD) constructed under the AIDA project (Advanced European Infrastructures for Detectors at Accelerators) was used to test the readout system, electronics, DAQ and reconstruction software. For the test beam, twelve planes, consisting of 16 scintillator bars $10\times 10\times 1000$~mm$^3$, read out on both sides by S12571-025C Hamamatsu MPPCs  along alternating $x$ and $y$ directions were instrumented (a total of 384 MPPCs). The beam consisted of muons and pions from $\sim$200~MeV/c up to 10~GeV/c. Figure~\ref{fig:AIDA} shows a schematic of the TASD and figure~\ref{fig:beamProfile} shows the evolution of the beam profile as measured by the TASD during the test beam and reconstructed with the SaRoMan software.

The electronic readout consists of custom-made Front End Boards (FEB). Each FEB consists of three 32 channels CITIROC ASICs. Each channel has a 12-bits ADC reading out at 40 Mega-samples per second per channel. The data processing is carried out by an Altera ARIA5 FPGA that controls the 400 MHz timing. The slow control readout is carried out with a USB3 or a Gigabit RJ45 chain and there is also an externally propagated Trigger signal.

\section{Baby MIND software}
The software suite used for the Baby MIND detector is the Simulation And Reconstruction Of Muons And Neutrinos (SaRoMaN) software. SaRoMaN can handle simulated single particle data from GEANT4 \cite{Geant4}, simulated neutrino data from GENIE \cite{GENIE} as well as real beam data. The digitisation software converts $x$ and $y$ bar hits into three-dimensional space points and performs event building from the data acquisition. The generic detector model is handled using GDML (Geometry Description Markup Language) files \cite{GDML}, allowing to utilize the same geometry in both the simulation and the reconstruction. The reconstruction software uses RecPack, a Kalman filter fitting package used to improve trajectory fits by using preliminary fitting parameters and a geometry description to improve the trajectory parameters \cite{2012apsp.conf..954C}.

\begin{figure}[h!]
	\begin{minipage}{0.49\linewidth}
		\centerline{\includegraphics[width=0.9\linewidth]{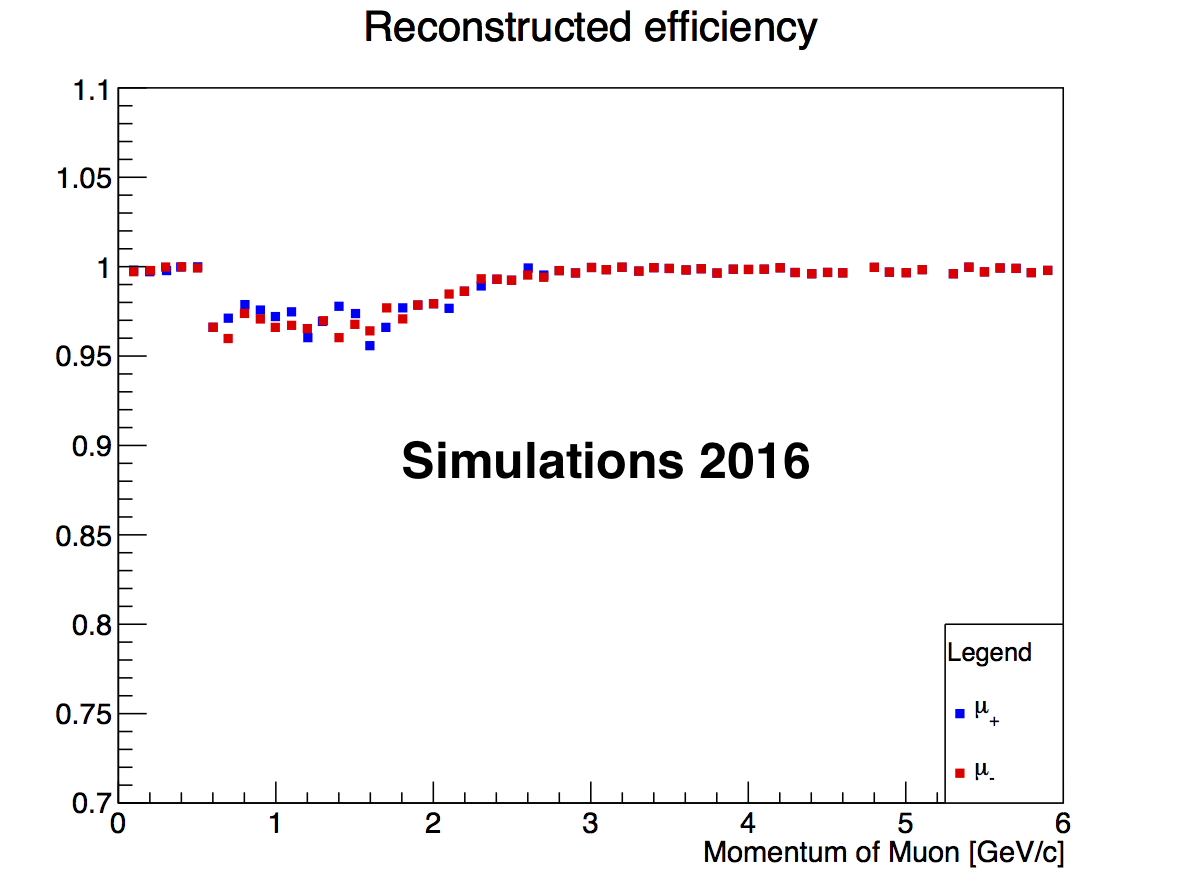}}
			\caption[]{ Reconstruction efficiency for a single muon pencil beam.}
		\label{fig:refEff}
	\end{minipage}
	\hfill
	\begin{minipage}{0.49\linewidth}
		\centerline{\includegraphics[width=0.9\textwidth]{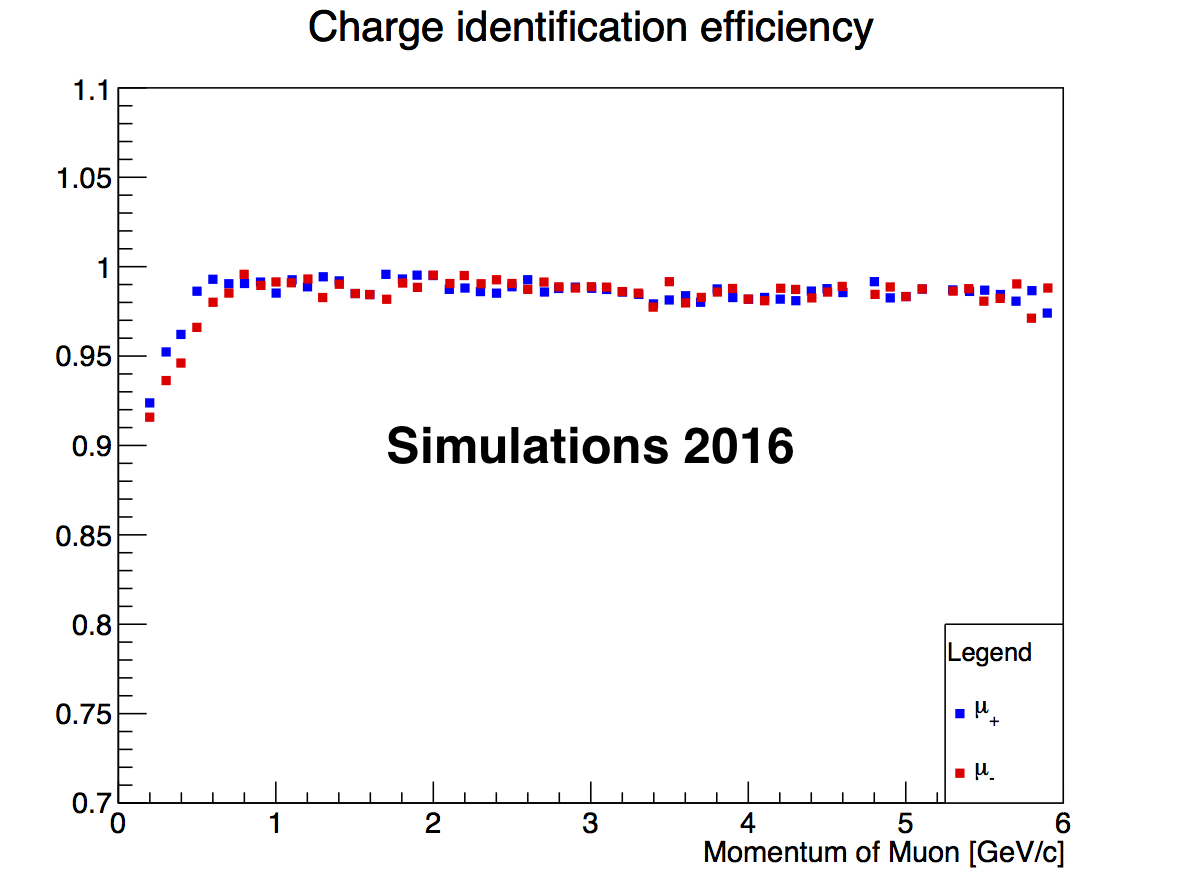}}
		\caption[]{Charge identification efficiency for a single muon pencil beam.}
		\label{fig:chargeID}
	\end{minipage}
\end{figure}

Figure~\ref{fig:refEff} shows the current reconstruction efficiency, using simulated data in Baby MIND, defined as the number of reconstructed tracks from the number of muons in the simulation.  For a single muon pencil beam parallel to the detector axis, the efficiency is more than 95\% for full range (0.2-6 GeV/c). Figure~\ref{fig:chargeID} shows the current charge identification efficiency, which is defined as the number of tracks with the correctly assigned charge out of all the reconstructed trajectories.  For a single muon pencil beam parallel to the detector axis, the efficiency is more than 90\% in the expected full reconstructible momentum range (0.2-6 GeV/c).

\section{Outlook}
The construction of the Baby MIND detector is ongoing and taking place at CERN. The magnetised steel plates are being assembled at CERN and the scintillator bars with wavelength shifting fibres are being constructed by INR Moscow. The detector construction and testing schedule is as follows:
\begin{itemize}
	\item The full Baby MIND detector will be finalised in May 2017;
	\item A test beam will be carried out in June 2017 to characterise the full Baby MIND detector;
	\item Shipping of the Baby MIND detector to Tokai, Japan in July 2017;
	\item Installation of the Baby MIND detector in the WAGASCI experiment in the Autumn 2017.
\end{itemize}


\if{0}

\fi

\bibliographystyle{unsrt}
\small
\bibliography{WAGASCI-bibliography}

\begin{thebibliography}{1}

\bibitem{NP05}
{CERN Neutrino Platform Project (NP05): the Baby MIND Magnetised Iron Neutrino
  Detector }.
\newblock {\em http://cenf-baby-mind.web.cern.ch}.

\bibitem{Koga:2015iqa}
T.~Koga et~al.
\newblock {Water/CH Neutrino Cross Section Measurement at J-PARC (WAGASCI
  Experiment)}.
\newblock {\em JPS Conf. Proc.}, 8:023003, 2015.

\bibitem{Noah:2015jgd}
Etam Noah.
\newblock {The WAGASCI experiment at JPARC to measure neutrino cross-sections
  on water}.
\newblock {\em PoS}, EPS-HEP2015:292, 2015.

\bibitem{Abe:2013hdq}
K.~Abe et~al.
\newblock {Observation of Electron Neutrino Appearance in a Muon Neutrino
  Beam}.
\newblock {\em Phys. Rev. Lett.}, 112:061802, 2014.

\bibitem{Geant4}
S.~Agostinelli et~al.
\newblock {Geant4—a simulation toolkit }.
\newblock {\em Nuclear Instruments and Methods in Physics Research Section A:
  Accelerators, Spectrometers, Detectors and Associated Equipment}, 506(3):250
  -- 303, 2003.

\bibitem{GENIE}
C.~Andreopoulos et~al.
\newblock {The GENIE Neutrino Monte Carlo Generator}.
\newblock {\em Nucl. Instrum. Meth.}, A614:87--104, 2010.

\bibitem{GDML}
R.~Chytracek, J.~Mccormick, W.~Pokorski, and G.~Santin.
\newblock Geometry description markup language for physics simulation and
  analysis applications.
\newblock {\em IEEE Transactions on Nuclear Science}, 53(5):2892--2896, Oct
  2006.

\bibitem{2012apsp.conf..954C}
A.~{Cervera-Villanueva}, J.~J. {G{\'o}mez-Cadenas}, and J.~A. {Hernando}.
\newblock {RecPack, a general reconstruction toolkit}.
\newblock In S.~Giani et~al., editors, {\em Astroparticle, Particle, Space
  Physics and Detectors For Physics Applications - Proceedings of the 13th
  ICATPP Conference}, pages 954--960. {World Scientific Publishing}, 2012.

\end{thebibliography}


\begin{thebibliography}{99}


@article{NP05,
Title = {{CERN Neutrino Platform Project (NP05): the Baby MIND Magnetised Iron Neutrino Detector }},
Journal = {http://cenf-baby-mind.web.cern.ch}
}

\bibitem{Mesmer}
F. A. Mesmer, Proc. Wien. Acad. Sci. {\bf 13}, 1564, 1593 (1762).

\bibitem{diCenzo}
A. L. di Cenzo, Trans. Acad. Ducal.  Milan., {\bf 23}, 2647 (1771).

\bibitem{Muller}
For an exhaustive review of the Prussian literature, see A. D. M\"uller,
in Workshop on Action at a Distance, F. Eisenschmidt, ed. (Springer,
Berlin, 1774).

\bibitem{daPonte}

L. da Ponte, Trans. N. Y. Acad. Sci., {\bf 3}, 27 (1795).


\end{thebibliography}
 
\end{document}